# An *in silico* drug repurposing pipeline to identify drugs with the potential to inhibit SARS-CoV-2 replication

Méabh MacMahon[1,2,†], Woochang Hwang[1,†], Soorin Yim[3], Eoghan MacMahon[4], Alexandre Abraham[5], Justin Barton[6], Mukunthan Tharmakulasingam[7], Paul Bilokon[8,9,10], Vasanthi Priyadarshini Gaddi[11], Namshik Han[1,12,*]

[1]Milner Therapeutics Institute, University of Cambridge, Cambridge, UK

[2]Centre for Therapeutics Discovery, LifeArc, Stevenage, UK

[3]Department of Bio and Brain Engineering, KAIST, Daejeon, Republic of Korea

[4]UCD School of Chemistry, University College Dublin, Dublin, Ireland

[5]Dataiku, Paris, France

[6]Institute of Structural and Molecular Biology, Birkbeck, University of London, London, UK

[7]Centre for Vision, Speech and Signal Processing, University of Surrey, Guildford, UK

[8]Department of Computing, Imperial College London, London, UK

[9]Department of Mathematics, Imperial College London, London, UK

[10]Thalesians Ltd, London, UK

[11]Centre for Genetics and Genomics Versus Arthritis, Division of Musculoskeletal and Dermatological Sciences, School of Biological Sciences, Faculty of Biology, Medicine and Health, The University of Manchester, Manchester, UK

[12]Cambridge Centre for AI in Medicine, Department of Applied Mathematics and Theoretical Physics, University of Cambridge, Cambridge, UK

[†]These authors contributed equally to this work and share first authorship

**\* Correspondence:**
Namshik Han
nh417@cam.ac.uk

**Keywords: COVID-19, Triamcinolone, Gallopamil, drug repurposing, artificial neural network, drug mechanism, virus replication**

**Abstract**

Drug repurposing provides an opportunity to redeploy drugs, which ideally are already approved for use in humans, for the treatment of other diseases. For example, the repurposing of dexamethasone and baricitinib has played a crucial role in saving patient lives during the ongoing SARS-CoV-2 pandemic. There remains a need to expand therapeutic approaches to prevent life-threatening complications in patients with COVID-19. Using an *in silico* approach based on structural similarity to drugs already in clinical trials for COVID-19, potential drugs were predicted for repurposing. For a subset of identified drugs with different targets to their corresponding COVID-19 clinical trial drug, a mechanism of action analysis was applied to establish whether they might have a role in inhibiting the replication of SARS-CoV-2. Of sixty drugs predicted in this study, two with the potential to inhibit SARS-CoV-2 replication were identified using mechanism of action analysis. Triamcinolone is a corticosteroid that is structurally similar to dexamethasone; gallopamil is a calcium channel blocker that is structurally similar to verapamil. In silico approaches indicate possible mechanisms of action for both drugs in inhibiting SARS-CoV-2 replication. The identification of these drugs as potentially useful for patients with COVID-19 who are at a higher risk of developing severe disease supports the use of in silico approaches to facilitate quick and cost-effective drug repurposing. Such drugs could expand the number of treatments available to patients who are not protected by vaccination.

**1 Introduction**

The novel virus, severe acute respiratory syndrome coronavirus 2 (SARS-CoV-2), which causes Coronavirus disease 2019 (COVID-19), was first identified in Wuhan in December 2019 and by January 2022 had spread to 192 countries or regions and infected over 335 million people, killing more than 5 million[1]. COVID-19 infection has a wide spectrum of symptoms. Those with mild infections may be asymptomatic, have anosmia[2] or mild respiratory symptoms[3], whereas severe infections lead to acute respiratory distress syndrome (ARDS) and potentially death[4]. There are certain populations known to be at higher risk of developing severe COVID-19, including people over 70 years old[5], those who are obese[6], diabetic[6], [7] or pregnant[8], or who are recipients of a solid organ transplant[9]. Despite several highly efficacious vaccines being deployed against SARS-CoV-2[10]–[13], the pandemic is still ongoing. Vaccine efficacy wanes with time[14] and this as well as breakthrough infections[15] and unequal vaccine distribution between higher and lower income regions[16] all contribute to the ongoing situation. Moreover, SARS-CoV-2 variants of concern[17]–[19] also alter the capacity to control infection rates[20] [21].

Current treatments for those infected with SARS-CoV-2 and at high risk of developing severe disease[22] are split into early and late therapies. Therapies for early stage disease Molnupiravir[23], [24] and nirmatrelvir (with ritonavir)[25], [26], which prevent replication of SARS-CoV-2, can be administered in patients with mild to moderate COVID-19 disease to prevent development of severe disease. Being orally administered, these drugs have an advantage over intravenously administered monoclonal antibodies[27], as they can be prescribed for use outside of a clinical setting and are less expensive[28], [29]. Hospitalised patients with severe disease can be treated with dexamethasone[30] and in the US, the Food and Drug Administration (FDA) has approved Veklury (remdesivir) for patients who are hospitalised with COVID-19[31].

Despite these advances in the treatment of mild and severe COVID-19, new drugs are still needed to treat this disease. For example, pregnant women are a high risk population for severe disease and neither nirmatrelvir or molnupiravir are currently approved for use in this population[32], [33]. There is also the risk of new variants arising that are resistant to approved drugs, as has happened with HIV mutations resistant to protease inhibitor treatment[34].

The development of a novel therapeutic can take many years[35] with high attrition rates, often owing to safety and toxicity concerns[36]. Drug repurposing aims to use existing approved or investigational drugs for therapeutic uses beyond the scope of the original medical indication[37]. This approach offers de-risked possibilities to identify safe, effective treatments faster and more economically than novel drug development[38]. Computational approaches are now routinely through multiple steps of the drug discovery process[39]–[44] and can be applied to repurposing of existing drugs[45], [46]. For example, network modelling [47]–[49], text mining[50] and various machine learning techniques[51]–[53] have been applied to repurposing drugs for COVID-19.

Chemical structure similarity is another approach that can be used in drug repurposing studies. It is based on the concept that structurally similar molecules often share similar biological function, and is a concept frequently used in drug discovery[54]. In this study, this approach has been used to identify drug candidates that could be repurposed to treat COVID-19 based on structural similarity to drugs already in clinical trials for COVID-19. A list of COVID-19 clinical trial drugs (C19-CTDs) was obtained from DrugBank[55]. Such C19-CTDs were used as a starting point because there are sufficient C19-CTDs with hypothesis-driven utility in COVID-19 and this also includes the few drugs that have been approved for the treatment of COVID-19. Where chemical similarity identified structurally similar drugs that have different targets to the C19-CTD, an additional network analysis was used to predict the probable mechanism of action and to determine whether utility in inhibiting SARS-CoV-2 replication was likely.

## 2 Results

### 2.1 Identification of drugs based on chemical structure similarity

To identify drugs that could be safely and rapidly deployed to treat COVID-19, the ChEMBL API was used to computationally predict structurally similar drugs (SSDs) to the C19-CTDs identified in DrugBank. Similarity was based on the Tanimoto coefficient (Tc), a similarity measure widely used in molecular fingerprint comparison[56]. When these analyses were initiated, there were 7,682 compounds listed in ChEMBL[57] as in clinical trials for any indication at Phase II or higher (including approved drugs). Among these 7,682 compounds, 101 compounds were determined to have structural similarity to 153 C19-CTDs, with a Tanimoto coefficient > 70 (Figure 1) .

### 2.2 Short-listing SSDs by step-by-step triage process

To eliminate any redundancy in the list of SSDs, several stages of filtering were applied (Figure 2). Initially, any SSDs which were already in the list of C19-CTDs were removed. For example, verapamil is an SSD because of its similarity to C19-CTD etripamil. However, verapamil is also a C19-CTD, so verapamil was removed. Next, any duplicated drugs were removed. For instance, taribavirin and taribavirin hydrochloride were both identified as similar to ribavirin (similarity score 76.2). However, since taribavirin hydrochloride is the acid salt of taribavirin, and once solubilized in the body (at approximately pH 7.4), both structures will exist in the same deprotonated form and have the same behaviour, these were considered as effective duplicates and the acid salt was removed from the list. After the removal of these duplicates, 65 unique SSDs remained. This list of 65 SSDs was compared to lists of drugs predicted from 6 other *in silico* repurposing papers which employed different techniques[47]–[52], 19 SSDs were predicted in other repurposing papers (Supplementary Table 1). These drugs were from 43 different Anatomical Therapeutic Chemical (ATC) classes (Supplementary

Table 1), with the top 3 most frequent classes being classes J01-"Antibacterials for systemic use", L01-"Antineoplastic agents" and S01-"Ophtamoligicals. Finally, 5 antineoplastic agents with the potential for severe side effects[58]–[60] (designated ATC code L01) were excluded from the list of SSDs, as they would be unsuitable to treat mild COVID-19 infections. This reduced the final list of SSDs to 60.

**2.3 Prioritisation of SSDs by their biological function**

For the identified SSDs which have similar chemical structure to C19-CTDs, the next stage of the analysis was to map their biological functions based on the proteins they target and the role of these proteins in the context of early SARS-CoV-2 infection. To do this, the SARS-CoV-2-induced protein (SIP) network[47] of interactions that occur at 24-hour after SARS-CoV-2 infection was used as the base network to investigate SSDs in the context of COVID-19. The 24-hour SIP network was used because a high concentration of the key proteins in this network are involved in viral replication and there is a significantly increased number of protein-protein interactions (PPI) in RNA- or viral-replication-related pathways[47], which this study wanted to examine in greater detail. This SIP network contains a computationally constructed PPI layer that connects directly interacting proteins (DIPs) and differentially expressed proteins (DEPs) identified by mass spectrometry[47], [61], [62].

To prioritise SSDs by their biological function, the targets of SSDs and C19-CTDs were compared. The comparison categorised the targets into three groups (Supplementary Table 1). In the first group, 16 SSDs among 60 short-listed SSDs have identical targets to their corresponding C19-CTD and would therefore have the same biological function. In the second group, 11 SSDs have the same targets as their corresponding C19-CTD in the context of proteins affected in early SARS-CoV-2 infection (ie. proteins within the 24-hour SIP network) and would therefore have the same biological function in this context. In the third group, 17 SSDs have some different targets from their corresponding C19-CTD, indicating they might have different biological function. Additionally, 16 SSDs were excluded due to lack of target information meaning no comparison could be carried out.

Since the third group of 17 SSDs target different proteins, they cannot be assumed to have the exact same biological implications as the C19-CTD from which they were predicted. This group might also have an advantage in that SSDs that have different biological implication to their C19-CTD could be particularly appropriate as drug repurposing candidates: if they have utility despite a different indication to the C19-CTD then treatment options available to patients could be expanded. For this reason, SSDs in this group underwent further analysis establish their MoA in the context of early SARS-CoV-2 infection.

**2.4 Mechanism of action analysis of SSDs in relation to COVID-19**

An initial pathway enrichment analysis carried out on the target proteins of all 28 drugs (17 SSDs and 11 C19-CTDs) identified a set of 488 biological pathways. As a measure of the extent to which the targets of each drug were associated with any of these 488 pathways, an F1 score was calculated for each drug-pathway association. From these, an F1 score matrix was generated, with 488 F1 scores for each of the 28 drugs, scoring their association with each of the 488 identified pathways.

The unsupervised training of a Self-Organizing Map (SOM) with this F1 score matrix generated 28 SOM component plane heatmaps (Supplementary Figure 1; Figure *3*). Each heatmap represents one drug, and each hexagon in the heatmap represents unique pathways which have clustered together, and which have the same position across all heatmaps (corresponding to drugs) (Figure *3*B-F). As the position of hexagons remains consistent across all SOM heatmaps, this allows direct visual comparison between pathways affected by each drug. To check how similar drug-pathway associations were between SSDs and their corresponding C19-CTD, the correlation coefficient (r) between SOM heatmap values for each pair was calculated (Supplementary Table 2). This was carried out to narrow the focus on SSDs with different biological implications to their corresponding C19-CTDs. 12 SSDs which were not correlated with their corresponding 8 C19-CTDs (r < 0.5) were identified (Figure 3A). These SSDs are triamcinolone and beclomethasone (which are similar to dexamethasone), Omega-3-carboxylic acids (which is similar to icosapent), beta-carotene (which is similar to isotretinoin), dextrothyroxine (which is similar to Liothyronine), xylitol (which is similar to mannitol), Clinidipine (which is similar to nimodipine), gallopamil (which is similar to verapamil) and four SSDs similar to zinc sulfate.

### 2.5 *in silico* target validation of SSDs which have potential in treating early COVID-19

It was reasoned that drugs which target viral replication might be more useful for the treatment of early-stage disease, while drugs whose targets are linked with immune-response-associated pathways might be better suited for later-stage COVID-19 infection. To assess whether any of the 12 SSDs would be useful in treating early COVID-19 infection the capacity of these drugs to impact pathways related to viral replication was investigated using the SOM generated data (Figure *3*F). For example, the correlation coefficient between dexamethasone and both SSDs predicted from it (beclomethasone and triamcinolone) was lower than 0.5, making it probable that these drugs have an MoA that is distinct from dexamethasone (Figure *3*A1, Supplementary Table 3). The SOM heatmap for beclomethasone is associated with pathways related to the immune response, including the estrogen receptor (ESR)[63], glycosylation[64], mitogen activated protein kinase (MAPK), activation protein 1[65] (AP-1) and phospholipase A2 (PLA2)[66] pathways (Figure *3*F). Interestingly, triamcinolone is associated with the high-density lipoprotein (HDL) remodelling pathway (Figure *3*F), a pathway that is reported to be strongly associated with coronavirus replication [67], [68].

Using the same approach, gallopamil, omega-3-carboxylic acids and beta-carotene were also identified for treating COVID-19 at an early stage of infection. The SOM heatmaps for gallopamil and triamcinolone are highly similar (Figure *3*A1 and Figure *3*A7), and unsurprisingly gallopamil is also associated with HDL remodelling (Figure *3*F). Omega-3-carboxylic acids is associated with fatty acid beta-oxidation, which is related to virus replication[69]. Beta-carotene is associated with the retinoic acid (RA) pathway, which is also related to virus replication. However, omega-3-carboxylic acids and beta-carotene are both listed as discontinued in the FDA approved drugs database[70] and so they were not analysed further in relation to early COVID-19. Although beta-carotene appears to be available in other jurisdictions, it is linked with increased risk of lung cancer[71], [72] and possibly pneumonia[73], [74] in individuals who smoke. Smoking is associated with increased severity of disease and death in hospitalized COVID-19 patients[75].

As a result of these analyses, only triamcinolone and gallopamil were analysed further for a potential impact in patients with early and mild to moderate COVID-19 symptoms. Triamcinolone is a corticosteroid, which is approved for the treatment of a range of inflammatory conditions including arthritis, asthma, and skin conditions. Triamcinolone inhibits nuclear factor kappa-B, which decreases the production of pro-inflammatory signals such as interleukin-6 (IL-6), interleukin-8, and monocyte chemoattractant protein-1[76]. IL-6 has an important role in cytokine release syndrome[77], which is triggered in patients with severe COVID-19 symptoms. Sustained elevation of IL-6 is also linked to death in acute respiratory distress syndrome[78] (ARDS), the respiratory condition experienced by critically ill patients with COVID-19.

Gallopamil is an L-type calcium channel blocker and an analogue of verapamil. Gallopamil is not yet approved for the treatment of any disease, although it has been tested in a Phase II trial as an orally administered drug for patients with severe asthma[79]. Gallopamil targets membrane metalloendopeptidase (MME), which is a type 2 transmembrane glycoprotein that cleaves angiotensin converting enzyme 1 (ACE) and ACE2. Since ACE2 is a cellular entry point for SARS-CoV-2 into human cells[80], this could be a potential mechanism through which gallopamil impacts SARS-CoV-2 replication. Additionally, ACE directly interacts with the HDL remodelling pathway proteins APOA1, APOE and ALB, as well as kinases that are activated in response to viral infection (AKT1, MAPK3, and MAPK1).

**2.6 Triamcinolone and gallopamil targets impact SARS-CoV-2 replication**

To establish the potential effect of triamcinolone and gallopamil on the replication of SARS-CoV-2, *in silico* validation was performed on their targets within the SIP network[47]. The proximity of the targets of each drug to a group of kinases that are predicted[81] to be active in the first 24-hours post infection with SARS-CoV-2 was measured. Both triamcinolone and gallopamil have 3 targets in the SIP network, and for both drugs, a permutation test showed that their targets are significantly ($P <0.05$) closer to the active kinases identified by Bouhaddou *et al*.[81] than 10,000 randomly selected groups of three proteins (Figure 4A). Indeed, both drugs directly interact with some of these active kinases identified by Bouhaddou *et al*.[81] (Figure 4B).

As a further *in silico* check of the potential efficacy of triamcinolone and gallopamil in inhibiting SARS-CoV-2 replication, the activity levels (expression patterns) of the target genes of these two drugs in data from patients with moderate symptoms of COVID-19 was investigated. Viral replication is known to be active in patients with moderate COVID-19 symptoms but decreased in patients who are severely ill with COVID-19[82]. Using the SOM results, a list of the proteins and pathways that are targeted by the two drugs was generated (Figure 3F, Supplementary Figure 2). To examine pathways being targeted, drugs were mapped to key pathways using SOM component plane heatmap (Figure 3A) and pathway clusters (Figure 3F, Supplementary Figure 2). For example, in the plane of the SOM component of triamcinolone (Figure 3A), the yellow hexagons are the triamcinolone associated pathways. The positions of the yellow hexagons in the triamcinolone SOM component plane can be mapped to the HDL remodelling pathway cluster, whose location can be seen in Figure 3F. This cluster is also associated with gallopamil (Figure 3F) and transcriptome analysis was used to identify upregulated pathways in this cluster. The neighbouring proteins of the drug's target proteins within each pathway were examined; that is, the proteins within each pathway that directly interact with the target proteins in the SIP 24-hour network (Figure 4B). The proteins in the HDL remodelling pathway were found to be significantly upregulated compared with a random selection of the same number of proteins (the drugs target proteins plus neighbouring proteins of the HDL-remodelling pathway) (Mann Whitney U test p-value <0.05, Figure 4C, Supplementary Table 4). Of the targets of the two drugs, ACE, and serpin peptidase inhibitor clade A member 6 (SERPINA6) directly interacted with APOA1 in the SIP 24-hour network. APOA1, which is a major transporter of HDL-cholesterol and is essential for HDL metabolism and remodelling[83, p. 1], has been linked *in silico* to the severity of COVID-19[84], [85]. The subnetwork for each drug (Figure 4B) shows that both gallopamil and triamcinolone interacted with proteins that directly interact with SARS-CoV-2, with kinases proposed to be active during early infection, with the HDL remodelling pathway including APOA1, and with proteins that show expression changes in the first 24 hours of COVID-19 infection (Figure 4B, Supplementary Video 1). As the target proteins of gallopamil and triamcinolone directly interact with these key proteins, this data set indicates mechanisms by which these drugs might impact early COVID-19 infection.

## 3 Discussion

In this study, an *in silico* approach was taken to identify drugs that have structural similarity to drugs already in clinical trials for the treatment of COVID-19. Of 65 unique drugs predicted, 19 have also been predicted by other *in silico* repurposing studies which used different approaches to this study. This overlap with other repurposing studies suggests lends support to the validity of this approach.

Of the 65 drugs, 16 have a highly similar structure and the exact same targets, therefore should the drug to which they are similar succeed in clinical trials, these drugs would be expected to be effective in patients with COVID-19. 11 others have the same protein targets within a network of proteins that are predicted to be active in patients 24 hours after infection with SARS-CoV-2 and should their corresponding C19-CTD be approved in early or mild-moderate infection, these should similarly be repurposed at this stage of infection. This study has further examined drugs that have structural similarity to drugs already in COVID-19 clinical trials but, target different pathways. The data from this study indicate that two of these, triamcinolone and gallopamil, have a distinct MoA related to virus replication and therefore might be useful in the early treatment of patients infected with SARS-CoV-2 who are at high risk of developing severe COVID-19. This supports the finding in a study by Zhou et al[49] published before there was available interactome data for SARS-CoV-2, which examined network interactomes of other human coronaviruses including SARS-CoV and MERS-CoV, and with which this study had 3 overlapping drugs, that triamcinolone was found to be significantly associated with the MERS-CoV-host interactome. Additionally, Gallopamil is a calcium channel blocker and this data supports studies which suggest the use of calcium channel blockers has been associated with decreased mortality in patients hospitalised with COVID-19 [86]–[88].

This *in silico* approach is based on publicly available data from online databases such as STITCH[89] and ChEMBL[57], and studies published in response to the pandemic[47], [62], [81], [82]. The focus of this study on drugs that are in clinical trials for COVID-19 was a logical starting point given the urgent nature of the continuing COVID-19 pandemic but has some obvious limitations. Firstly, as the controversy surrounding hydroxychloroquine[90] highlights, a drug being in a COVID-19 clinical trial does not necessarily mean it is likely to be effective. Also, there are likely to be many relevant drugs that have been missed in this study as they are not structurally similar to drugs that have already been in or are in a clinical trial for the treatment of COVID-19. An additional limitation of this study is that owing to the chemical-structure-based approach the focus was on small molecules only and not biologics. Lastly, STITCH was among the databases used to obtain target information for comparing C19-CTDs to SSDs; while this is an important and useful resource, it does not take account of incomplete drug information or off-target effects. Finally, although *in silico* validation has indicated promise for triamcinolone and gallopamil, these results are based on a protein-protein interaction (PPI) network generated from cell-derived mass spectrometry data and a PPI resource[91], which lack directionality for the protein interactions[47] and wet lab follow-up studies in experimental models are required to verify these effects and support their clinical potential.

### 3.1 Conclusions

This *in silico* study has predicted drugs that have potential for repurposing for the treatment of COVID-19. Triamcinolone and gallopamil have been identified by this approach as drugs with potential for targeting the replication of SARS-CoV-2. Of the two, triamcinolone is already approved for other indications, so could be deployed quickly. Gallopamil has passed Phase II trials and could be quickly prioritised into COVID-19 trials. Both require further investigation in *in vitro* and *in vivo* models, as well as clinical trials. It is likely that further diseases of zoonotic origin will emerge in the future[92], and the scientific community must be prepared to respond rapidly to future pandemic threats. This kind of *in silico* approach allows for a quick and cost-effective approach to repurposing drugs, which could be deployed in such a situation.

## 4 Materials and Methods

### 4.1 Obtaining list of structurally similar drugs

A list of 184 COVID-19 clinical trial drugs (C19-CTDs) was obtained from DrugBank[55] (date accessed 25/05/2020). There were 153 small molecules (Supplementary Table 5) and 31 biotech drugs, as this analysis was based on chemical structure, biotech drugs were excluded. Drugbank is an online bioinformatics and chemoinformatics database which contains data on drugs and their targets, for instance the structure of a drug and any clinical trials it has been involved in as well as approval status. ChEMBL is a database containing bioactivity data for drugs and drug-like molecules and tools for mining this information, including by searching for similar chemical structures within the ChEMBL database itself[57]. The ChEMBL application programming interface (API) for chemical structural similarity accepts input in the simplified molecular-input line-entry system (SMILES) format, which is a format for represents chemical structures using ascii strings[93]. For each C19-CTD, the structure in SMILES format was extracted from the DrugBank data (script in code repository). The ChEMBL API library (written in python) was used to obtain a list of structurally similar drugs (SSDs) which were phase II clinical trial or higher from ChEMBL (script in code repository). SSDs that were in Phase II clinical trials or higher for any condition were retained at this stage because they passed Phase I trials and have therefore been tested for safety in humans. The ChEMBL structural similarity tool used is based on the Tanimoto coefficient (Tc), a similarity measure widely used in molecular fingerprint comparison[56]

### 4.2 Filtering the list of SSDs

The list of SSDs was refined to remove drugs which fell into several categories (Figure 2). Tc > 70 was used as a similarity cut off. For context, a threshold of 70-75 is typical for structure based screening[94]. The list of SSDs was cleaned to remove drugs which are synonyms of C19-CTDs (using synonym information from Drugbank) and effective duplicates that behave the same in the body despite of a slightly different chemical structure (using salt information from Drugbank).

### 4.3 Obtaining target information for drugs

Target information for all drugs was collected from target information in DrugBank[55], targets in Therapeutic Target Database (TTD)[95] and targets with interaction confidence score > 0.8 in drug to target interactions from STITCH (v 5.0)[96, p. 5]. TTD is a database containing target-specific drug binding information about known therapeutic targets from literature[95]. STITCH is a database containing scored interactions between proteins and small molecules which has evidence from experimental/biochemical data, association in curated databases and co-mention in abstracts from the scientific literature[89]. (A score of 0.7 is considered high confidence by STITCH[97, p. 4]).

### 4.4 Filtering of drug list based on target information

The SARS-CoV-2-induced protein (SIP) network at 24 hours generated by Han et al[47] was used as the base network to investigate the COVID-19 pathways on which these structurally similar drugs (SSDs) are acting. This SIP network is a network constructed in silico of the protein-protein interaction pathways which are activated at 24 hours post COVID-19 infection. The methods for construction of this network are described in the Han et al paper[47]. Briefly, this network contains the directly interacting proteins (DIP) of SARS-CoV-2 identified by mass spectrometry[62], differentially expressed proteins (DEP) after infection by SARS-CoV-2 identified by mass spectrometry[61], and a layer of connections between the two built by obtaining all shortest paths between DIP proteins and DEP proteins using known protein-protein interaction (PPI) data as described in the Han et al paper[47].

**4.5 Distance between targets of SSD and the targets of C19-CTD**

The distance dc(S, C) is defined between S, the set of target proteins of a SSD, and C, the set of target proteins of a C19-CTD, as the shortest path length d(s, c) between all pairs of nodes s ∈ S and c ∈ C in the SIP 24-hour network[47]. Closest distance measure (equation (1)) was used to calculate the distance between a SSD's target to its corresponding C19-CTD's target in the SIP 24-hour network because it showed the best performance in drug-drug relationship prediction in a study by Cheng et al[98]. If the average distance <1 between the targets of any SSD and its corresponding C19-CTD in the SIP 24-hour network[47] the two drugs don't have different targets in the network.

$$d_c(S,C) = \frac{1}{\|C\|}\sum_{c \in C} min_{s \in S} d(s,c) \qquad (1)$$

**4.6 Drug-pathway associations**

To identify key pathways that are significantly enriched in the proteins that are targeted by 28 SSDs and C19-CTDs, pathway enrichment analysis was conducted using R (v 3.5.2) package gprofiler2[99] gost function, which uses a hypergeometric test to identify genes overrepresented in known pathways for functional enrichment analysis of gene list (FDR-BH < 0.05). Reactome is a database of manually curated and peer reviewed biological interactions, including curated human pathways organised based on biological function[100]. Although there are a number of pathway databases[101], [102], Reactome (accessed on 15/05/2020) was chosen for pathway enrichment analysis because it is the most actively updated public database containing human pathways[103]. To understand the MoA of the 28 drugs, pathway enrichment analysis was performed on the targets of each of these drugs. The number of target proteins of the 28 drugs ranges from 2 to 125. The average number of target proteins of the 28 drugs is 14. Drugs targeting fewer than six proteins have no significantly enriched pathways. To overcome this problem if drugs have less than the average number of target proteins (ie. 14), neighbour proteins (ie. proteins with which they directly interact) of the target proteins in the SIP 24-hour network were included in their target list for enrichment analysis, to 14 proteins. Significantly enriched biological pathways of target proteins for each of the 28 drugs were integrated, resulting in 488 key pathways. The Reactome pathway has a hierarchical structure among pathways. The lower hierarchy pathway is more specific than the higher hierarchy pathway. The parent pathway semantically includes its child pathways. In the process of integrating the enriched pathways per drug, the lowest-possible hierarchy pathways were used to avoid overlapping biological meaning among the hierarchical pathways.

Based on these identifications, a matrix containing F1 scores of the 28 drugs and the 488 key pathways was generated for drug-pathway association. F1 score is the harmonic mean of precision and recall values in a classification and is an accuracy metric used here to calculate enrichment accuracy. The Reactome pathway enrichment analysis for the 28 drugs using gprofiler2[104] provides enrichment p-values, precision and recall information that were used to produce the F1 scores. The meaning of precision here is that the proportion of a drugs target proteins that are annotated to the pathway. The meaning of recall here is that the proportion of the pathway gene sets that a drugs target proteins are located in. Precision and recall were used to construct a matrix of F1 score (F1=2(precision x recall)/(precision + recall) from the pathway enrichment analysis.

### 4.7 SOM analysis of pathways impacted by targets of drugs

Self-Organizing Map (SOM) [103] was used to analyse the MoA of the candidate drugs based on pathways impacted by their targets in this analysis. SOM, a type of artificial neural network, is an unsupervised machine learning technique used for dimensional reduction and clustering. SOM additionally has an ability to describe and visualise high dimensional data. For this reason, it was useful to directly compare the SOM component heatmaps of the 28 drugs for COVID-19 with one another. SOM also has the advantage of dimensional reduction to allow a clearer clustering result. SOM was used to calculate the low-dimensional abstractions, which are then clustered using k-means. This two-phase approach increases the efficiency of k-means clustering with a relatively small number of samples, a known limitation in hierarchical clustering algorithms including k-means[105]. Another advantage of SOM is noise reduction because the SOM abstractions are less sensitive to random variations than the input data. In addition, SOM offers a systematic arrangement of the candidate drugs to each neuron (or node) and hence to pathway clusters (Fig 3C).

The data used in training was the F1 score matrix for Drug-Pathway associations (key pathways by candidate drugs). From the SOM training results a Unified distance matrix is generated which captures the distribution of the trained artificial neurons in the data space, this contains the vector norms between the neighbouring SOM nodes and shows data density in input space. Each sub-unit is coloured according to distance between corresponding data vectors of neighbour units. Low distances areas (dark blue) have high data density (clusters) (Fig. 3B). Davies-Bouldin (DB) index [106], a method for identifying the optimal k number for k-means clustering ,was calculated based on the U-matrix to determine the optimal number of clusters. DB index is calculated by obtaining the average similarity of each cluster with the cluster most similar to it. The lower the DB index is, the better separated clusters are. The lowest DB index value occurred at 10 clusters (Fig. 3F). The k-means algorithm was then used to cluster pathways (Fig. 3F). The SOM component maps of key pathways were analysed based on the k-means clustering result (Fig. 3F) and mapped into two MoA categories based on the biological functions (Fig. 3F). The mapping result of key pathways to clusters is available in Supplementary Table 3. The SOM Toolbox package [107] for MatLab was used for this analysis with default settings and parameters.

### 4.8 In silico mechanism of action analysis of Triamcinolone and Gallopamil

To identify MoA for triamcinolone and gallopamil in COVID-19 their relationships within the SIP 24-hour SARS-CoV-2 network were identified. There are 14,827 proteins and 528,969 interactions in this network. Bouhaddou et al[81] predicted kinase activities in the first 24h post infection with SARS-CoV-2 based on the regulation of their known substates. From their data all kinases with activity absolute LogFC > 1.5 at any time point in the first 24h post infection for verification were used as these are being activated in early SARS-CoV-2 infection. To estimate the proximity of the targets of the drugs of interest to these active kinases in the SIP network, the distance of each target to each kinase was calculated and the average of these distances was taken. To establish significance, a permutation test was used as follows; each drug had 3 targets in the network, so 10,000 groups of 3 proteins present in the network were randomly generated and the average distance of these from the active kinases was calculated (Figure 4A, Supplementary Video 1). A permutation test is an exact test to establish statistical significance of a null hypothesis by randomly resampling all data points[108], it has been applied in network analysis based drug repurposing[47], [49], [98]. The closest 5% distances to the kinases were used as a cut off to establish significance of $P < 0.05$. To visualise the target proteins interacting with these kinases and other key COVID-10 proteins, subnetworks were plotted using Virtualitics Immersive Platform[109]. An additional movie file shows this in more detail (Supplementary Video 1)

**4.9 Expression analysis of the drugs target proteins using COVID-19 patients' data**

To indirectly validate the effects of triamcinolone and gallopamil in COVID-19, it was established whether the target proteins of these drugs are significantly changed between patients with moderate and severe COVID-19 as compared to the expression of random proteins. Arunnachalam et al.[82] provide Log2 fold change gene expression values between 4 COVID-19 moderate patients compared to 12 COVID-19 severe patients. In the boxplot shown in Figure 4C Expression of proteins belonging to pathways in the HDL remodelling cluster (Figure 3F) which were also among the neighbour proteins of each of the drugs targets in the SIP 24-hour network were compared to the same number of randomly selected control proteins, comparisons for the significantly upregulated pathway are shown in Figure 4C. To ensure the control proteins were unbiased, this random choice was repeated for 100 permutations to create control protein expression values. For example, if five of the target proteins and their neighbour proteins belong to the pathway in the HDL remodelling cluster, 500 proteins were randomly selected and made into control proteins. The Mann Whitney U test, a non-parametric version of the independent samples t-test[110], was performed to determine whether the expression two groups were significantly different between moderate and severe COVID-19 patients.

**5 Conflict of Interest**

MM is an employee of LifeArc. NH is a cofounder of KURE.ai and CardiaTec Biosciences and an advisor at Biorelate, Promatix, Standigm and VeraVerse. SY, EM, AA, JB, MT, PB and VPG declare no competing interests.

**6 Author Contributions**

Conceptualization, M.M., W.H. and N.H.; Methodology, M.M., W.H. and N.H.; Software, M.M., W.H. S.Y., M.T. and N.H.; Validation, M.M., W.H. and N.H.; Formal Analysis, M.M., W.H. and N.H.; Investigation, M.M, W.H., N.H., S.Y., E.M., A.A., J.B., M.T., P.B., and, V.P.G. ; Resources, N.H.; Data Curation, M.M., W.H., E.M. and N.H.; Writing – Original Draft M.M. and W.H.; Writing-Review and editing M.M, W.H., N.H., S.Y., E.M., A.A., J.B., M.T., and, V.P.G.;Visualization, M.M., W.H., E.M., and N.H; Supervision, N.H.; Project Administration M.M., W.H. and N.H. All authors read and approved the final manuscript

## 7 Funding

NH and WH are funded by Lifearc. SY is funded by Bio-Synergy Research Project (2012M3A9C4048758) of the Ministry of Science and ICT through the National Research Foundation.

## 8 Acknowledgments

The authors acknowledge dataswift Hack from home hackathon which brought them together on this project and Dr Daniel Baxter, Dr Zhi Yao, Dr Nicola McCarthy, Dr Rebecca Harris, Dr Alison Schuldt, Dr Kathryn Chapman, and Dr Tony Kouzarides for useful discussion and comments on manuscript.

## 9 Data Availability Statement

This study uses publicly available data from online databases and other research papers as cited in the text. No new data were generated. Further information and requests for resources and materials should be directed to and will be fulfilled by the lead contact, Namshik Han (nh417@cam.ac.uk). Code is available at the GitHub repository https://github.com/wchwang/COVID-19.HfH .

## 11 Figures

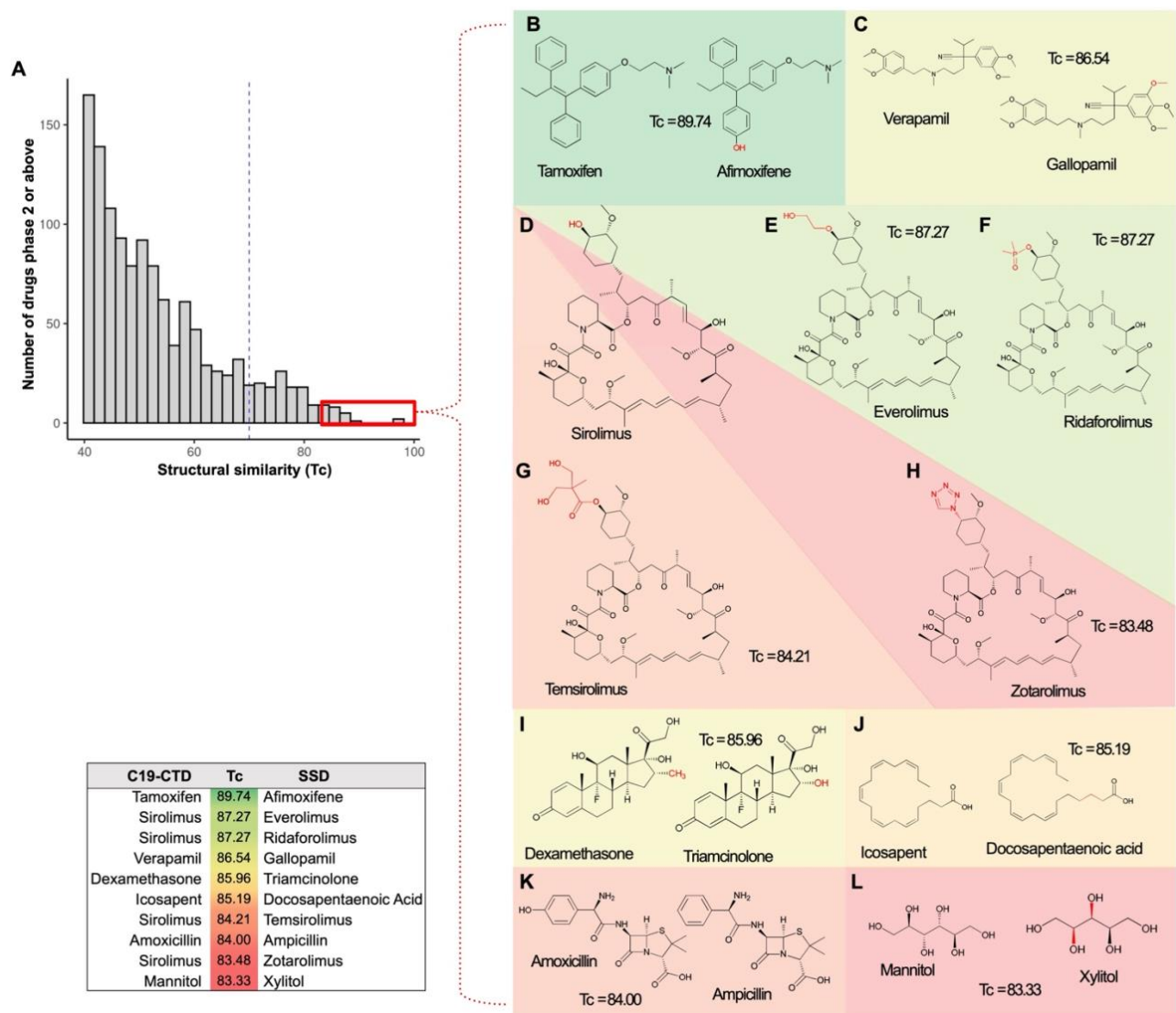

**Figure 1. Drug structural similarity screening** (A) Histogram showing the number of drugs in Phase II+ clinical trials with a Tc > 40 and < 100 that are similar to C19-CTDs. The cut off for similarity of 70 used in this study is shown by the blue dashed line. The red box highlights the drug structures shown in panels B-L in the form of a heatmap of similarity. (B) Tamoxifen predicts afimoxifene. (C) Verapamil predicts gallopamil. (D) Sirolimus predicts (E) everolimus, (F) ridaforolimus, (G) temirolimus, and (H) zotarolimus. (I) Dexamethasone predicts triamcinolone. (J) Icosapent predicts docosapentaenoic acid. (K) Amoxicillin predicts ampicillin. (L) Mannitol predicts xylitol. The structural differences are highlighted in red.

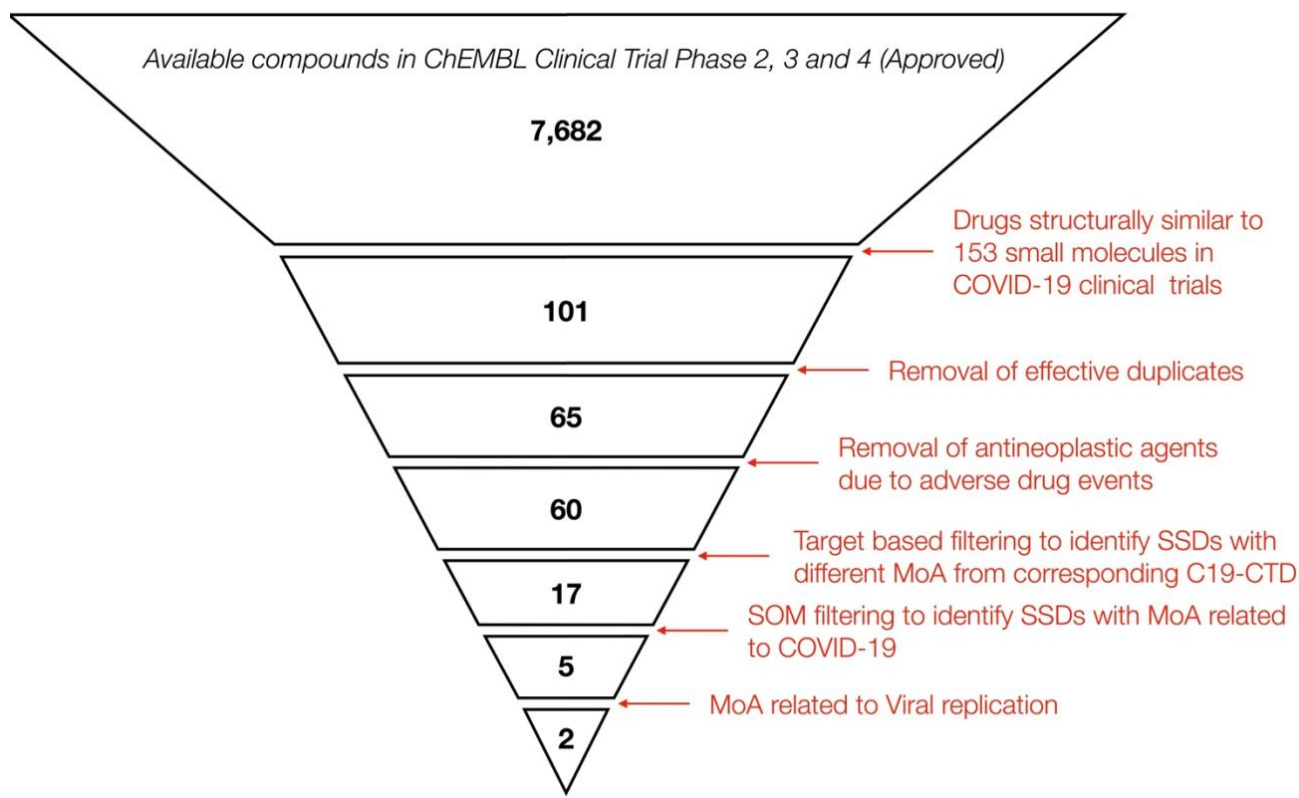

**Figure 2. Overview of SSD prediction and filtering.** This figure shows the workflow used to identify SSDs from C19-CTDs and filtering steps used to and identify those with potential for treating early COVID-19 including the number of SSDs retained after each filtering step.

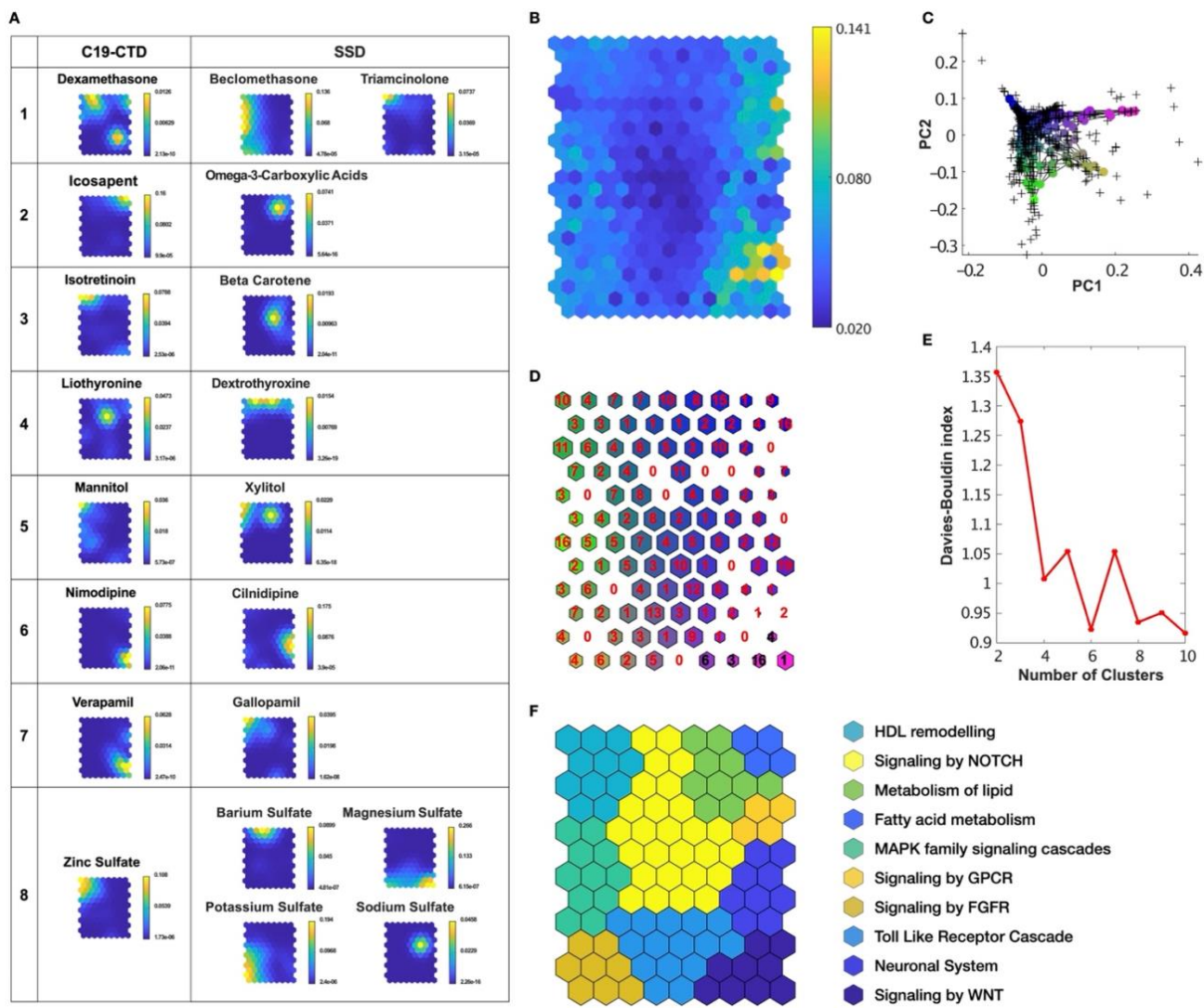

**Figure 3. Drug-pathway association study** (A) 20 SOM component plane heatmaps, one for each set of SSDs which were not correlated to their corresponding C19-CTDs. In each plane heatmap, the hexagon in a certain position corresponds to a set of unique pathways which has the same position across all heatmaps (drugs). Each hexagon is coloured according to the distance between corresponding data vectors of neighbour neurons, with low distances areas (bright yellow) indicating high data density. The 488 pathways assigned to hexagons in the SOM heatmap were clustered using a unified distance matrix (U-matrix) (B), that captures the distribution of the trained artificial neurons in the data space (C). The SOM also provided the clustering information based on the U-matrix and number of hits (pathways) per hexagon (neuron) (D). In addition, the Davies–Bouldin index (DBI) was used to establish the optimal number of clusters for *k*-means clustering (E). Using this information, 488 pathways were classified into 10 pathway clusters (F).

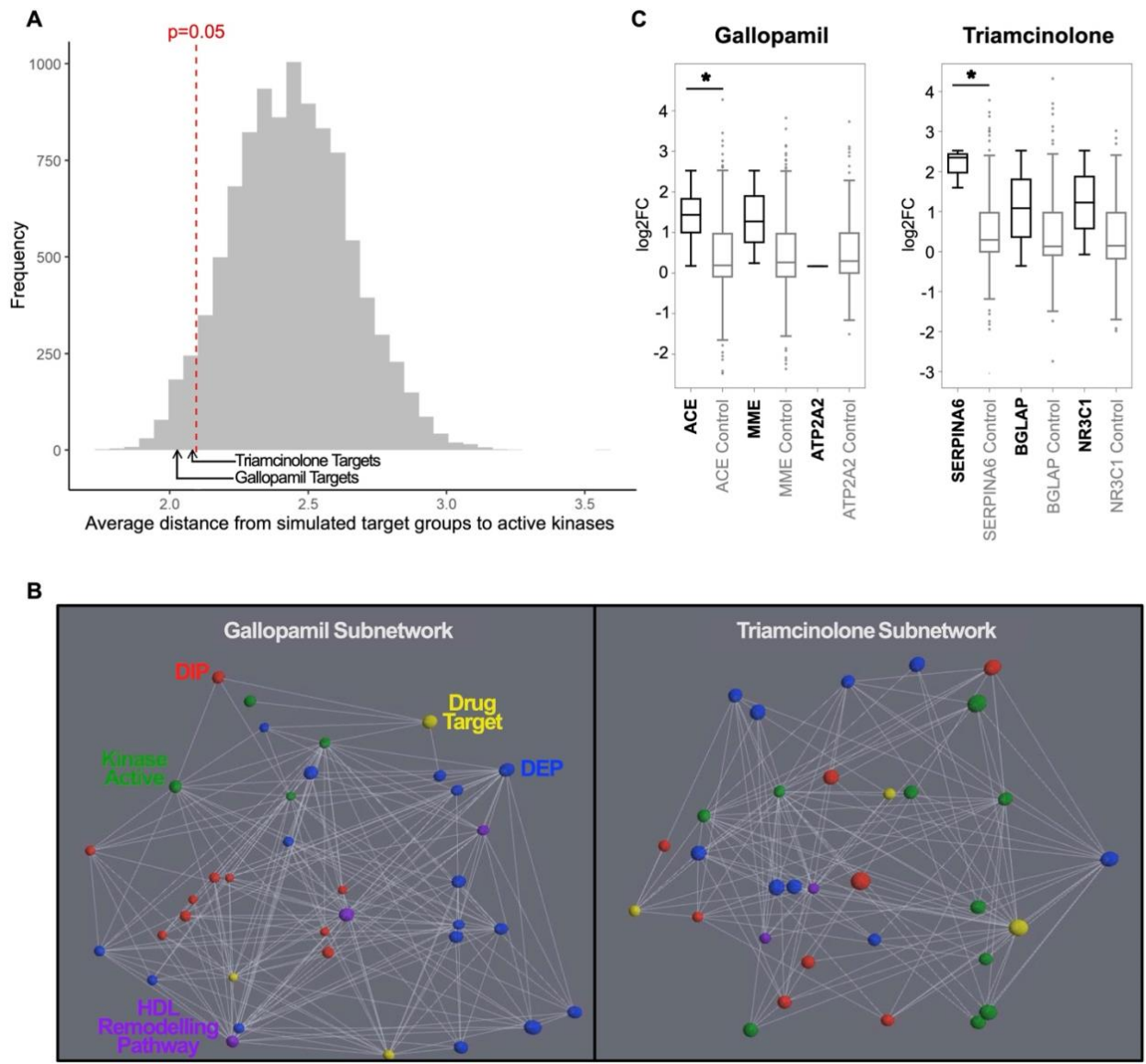

**Figure 4. *In silico* validation of gallopamil and triamcinolone.** (A) Histogram of the results of a permutation test, the average distances between 10,000 groups of 3 proteins and kinases active at 24 hours after infection with SARS- COV-2. The distances of the gallopamil and triamcinolone targets are indicated with arrows, and the red dashed line shows the permutation cut-off for the 5% of protein groups closest to the kinases (p-value = 0.05) (B) Subnetwork of gallopamil and triamcinolone target proteins and their direct interactors showing how the targets of both drugs interact with proteins, which are proposed to be important in COVID-19 infection. The targets of both drugs (yellow) interact with DIPs (red), DEPs (blue), active kinases (green), and the proteins related to the HDL remodelling pathway (purple). (C) Boxplots showing targets of drugs in HDL remodelling pathway and their neighbouring proteins are up-regulated compared to control in patients with moderate COVID-19 symptoms. The control results in the boxplot show fold-change of random genes with the same number of target proteins and neighbour proteins as the drugs of interest. Significance was tested by the Mann Whitney U test (p-value <0.05).